\documentclass[prl,aps,floats,twocolumn]{revtex4}%
\usepackage{amsmath}
\usepackage{amsfonts}
\usepackage{amssymb}
\usepackage{graphicx}%
\begin{document}
\title{Self-generated electronic heterogeneity and quantum glassiness \\in the high temperature superconductors}
\author{C. Panagopoulos}
\affiliation{Cavendish Laboratory and IRC in Superconductivity, University of Cambridge,
Cambridge CB3 0HE, United Kingdom}
\author{V. Dobrosavljevi\'{c}}
\affiliation{Department of Physics and National High Magnetic
Field Laboratory, Florida State University, Tallahassee, FL 32306}

\begin{abstract}
We present a systematic study of the spin and charge dynamics of
copper oxide superconductors as a function of carrier
concentration $x$. Our results portray a coherent physical
picture, which reveals a quantum critical point at optimum doping
($x=x_{opt}$), and the formation of an inhomogeneous glassy state
at $x < x_{opt}$. This mechanism is argued to arise as an
intrinsic property of doped Mott insulators, and therefore to be
largely independent of material quality and level of disorder.

\end{abstract}
\maketitle

Many interesting materials ranging from magneto-restrictive
manganite films \cite{dagotto-book} and field-effect transistors
\cite{abrahams-rmp01,bogdanovich-prl02}, to unconventional low
dimensional superconductors \cite{kivelson-rmp03}, find themselves
close to the metal-insulator transition. In this regime,
competition between several distinct ground states
\cite{kivelson-rmp03} produces unusual behavior, displaying
striking similarities in a number of different systems. Electronic
heterogeneity \cite{schmalian-prl00,dagotto-book} emerges, giving
rise to "mesoscopic" coexistence of different ordered phases.
Typically, a large number of possible configurations of these
local regions have comparable energies, resulting in slow
relaxation, aging, and other signatures of glassy systems. Because
the stability of such ordering is controlled by doping-dependent
quantum fluctuations \cite{pastor-prl99,mitglass-prl03} introduced
by itinerant carriers, these systems can be regarded as
prototypical quantum glasses - a new paradigm of strongly
correlated matter.

In this Letter we report the emergence and evolution of dynamical
heterogeneity and glassy behavior across the phase diagram of the
high-transition-temperature ($T_{c}$) superconductors (HTS). Based
on data of the spin and charge dynamics, we draw a phase diagram
(Fig. 1) and propose that self generated glassiness
\cite{schmalian-prl00} may be a key feature necessary to
understand many of the unconventional properties of both the
superconducting and the normal state.

\emph{Glassiness in the pseudo-gap phase.} In the archetypal HTS, La$_{2-x}%
$Sr$_{x}$CuO$_{4}$ (LSCO) the parent 2D antiferromagnetic
insulator (AFI) La$_{2}$CuO$_{4}$ displays a sharp peak in the
magnetic susceptibility at the Neel temperature $T_{N}=300K$.
$T_{N}$ decreases with hole-doping and the transition width
broadens (Fig. 2 -- upper panel). Concurrently, there is
systematic experimental evidence from various techniques and on
several HTS that a second freezing transition ($T_{F}$) emerges at
lower temperatures with the first added holes (Fig. 2)
\cite{niedermayer-prl98,panagopoulos-ssc03,matsuda-prb02,sanna-lanl04,ishida-prl04,kohsaka-prl04}.
At $x>0.02 $, $T_{N}=0$ but the short range order persists
\cite{panagopoulos-prb02}: the low-field susceptibility displays a
cusp at low temperatures and a thermal hysteresis below,
characteristic of a spin glass transition ($T_{g}$) (Fig. 2 --
upper panel). At $T<T_{g}$ the material
displays memory effects like \textquotedblleft traditional\textquotedblright%
\ spin glasses and is described by an Edwards-Anderson order
parameter \cite{chou-prl95}. Interestingly, it is at this doping
range that a pseudogap phase develops
\cite{kivelson-rmp03,hanaguri-nature04}.
 \begin{figure}[h]
\begin{center}
\includegraphics[
height=3.1872in, width=3.208in ]{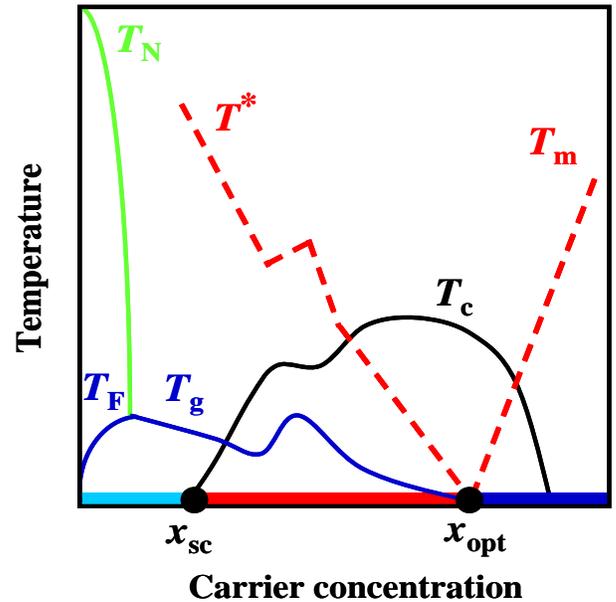}
\end{center}
\caption{Schematic plot indicating the three major ground state
regimes in the phase diagram of the archetypal HTS. $T_{N}$ is the
Neel temperature, $T_{F}$ and $T_{g}$ the onset of short range
freezing to an electronic glass, and $T_{c}$ the superconducting
transition temperature. At $x<x_{sc}$ the material is a glassy
insulator. At $x_{sc}<x<x_{opt}$ a microscopically inhomogeneous
conducting glassy state emerges, with intercalated superconducting
and magnetic regions. At $x=x_{opt}$ the system experiences a
quantum glass transition and at $x>x_{opt}$ the material
transforms into a homogeneous metal with BCS-like superconducting
properties. The superfluid density is maximum at $x=x_{opt}$. The
crossover scales $T^{*}$ and $T_{m}$ characterizing normal-state
transport (see text for details), vanish at the quantum glass
transition.}
\end{figure}\pagebreak

 The emergence of
electronic heterogeneity with charge doping is not unique to HTS.
It has been demonstrated in semiconductors
\cite{bogdanovich-prl02}, ruthanates \cite{nakatsuji-prl04},
nickelates \cite{sarrao-lanl04}, other copper oxides
\cite{sarrao-lanl04,sasagawa-prb02}, and manganites
\cite{bhattacharya-lanl04}, where transport experiments indicate
that at least some glassy features originate from slow charge
dynamics. Further evidence supporting that glassiness in the
charge and the spin channels emerge hand-in-hand was
recently provided by measurements of the dielectric constant on $La_{2}%
Cu_{1-x}Li_{x}O_{4}$ and $La_{2-x}Sr_{x}NiO_{4}$
\cite{sarrao-lanl04}. These materials are not superconducting, but
their spin response is almost identical to that of cuprate
superconductors - while the dielectric response is remarkably
similar to conventional (structural) glasses.

The glassy signatures in the spin channel of HTS suggest analogous
effects in the charge response, which we examine by measuring the
in-plane electrical resistivity, $\rho_{ab}$, of LSCO single
crystals (Fig. 2 -- lower panel, inset). The data for
$x$=0.01-0.04 show a crossover (resistivity minima) in
$\rho_{ab}(T)$ from metallic to insulating-like at a
characteristic temperature $T^{\ast}$ (Fig. 2 -- upper panel).
Although the crossover takes place over a wide temperature region
it clearly occurs at $T\ll T_{N}$ for $x<0.02$ and has a doping
dependence similar to $T_{F}$ and $T_{g}$. This similarity
indicates the association between short range order and charge
retardation. Moreover, the emergence of a glass order with the
first added carriers speaks against impurity effects but instead
for an intrinsic property.
\begin{figure}[h]
\begin{center}
\includegraphics[
height=5.4172in, width=3.1in ]{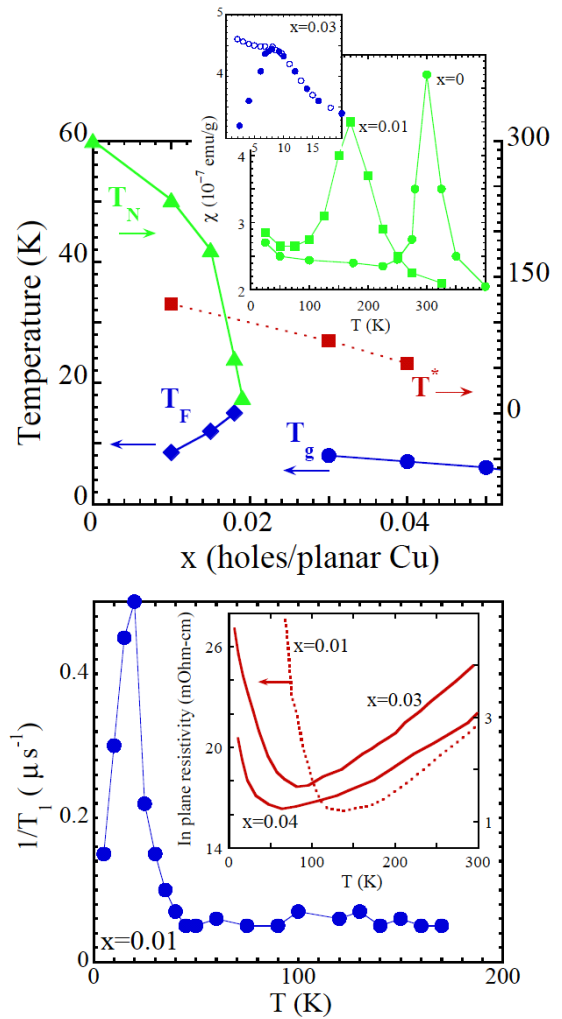}
\end{center}
\caption{The upper panel depicts the doping dependence of the Neel
temperature $T_{N}$, the second freezing $T_{F}$ and glass
temperature $T_{g} $. The latter two scales are determined by a
peak in the spin lattice relaxation $1/T_{1}$ (see $e.g.$, lower
panel) - data obtained by zero field $\mu SR$ on LSCO ($x=0.01$).
$T^{\ast}$ is a crossover from metallic-like to insulating-like
resistivity determined from the in-plane resistivity measurements
shown as inset to the lower panel. The inset in the upper panel
depicts the temperature dependence of the susceptibility for LSCO
($x=0,0.01$) single crystals with $H//c$. The associated inset
is data for $x=0.03$
showing the transformation of the material to a glass. }%
\label{fig1}%
\end{figure}

\emph{Coexistence of glassiness and superconductivity.} In light
of many unconventional properties of HTS below optimum doping
\cite{kivelson-rmp03}, it is important to probe the possible
correlation between the identified dynamical heterogeneities and
superconductivity (present for $x>x_{sc}=0.05$ for LSCO). Muon
spin relaxation ($\mu$SR) has been successful in identifying the
freezing of electronic moments under the superconducting dome of
various HTS
\cite{niedermayer-prl98,panagopoulos-ssc03,panagopoulos-prb02,kanigel-prl02,sanna-lanl04,hanaguri-GRC04}.
Figure 3 (inset) shows a typical example (LSCO, $x=0.08$
($T_{c}=21$ K)) of spectra with a glass transition at low
temperatures displaying an initial rapid relaxation. The amplitude
of the muon spin polarization reveals that all muons inside the
sample experience a non-zero local field indicating that the
magnetism persists throughout the entire volume of the sample.
However, the absence of a dip in the polarization function at the
lowest times indicates the presence of a large number of low field
sites -- superconducting and magnetic regions intercalated on a
microscopic ($\leq2$ nm) scale
\cite{niedermayer-prl98,panagopoulos-ssc03,panagopoulos-prb02,kanigel-prl02}.
These observations suggest the presence of magnetic
stripes/droplets, in agreement with independent indications
\cite{kivelson-rmp03,niedermayer-prl98,panagopoulos-ssc03,matsuda-prb02,sanna-lanl04,kohsaka-prl04,panagopoulos-prb02,kanigel-prl02}
from transport and spectroscopic studies for the
non-superconducting dopings.
\begin{figure}[ptb]
\begin{center}
\includegraphics[
height=2.9655in,
width=3.208in
]{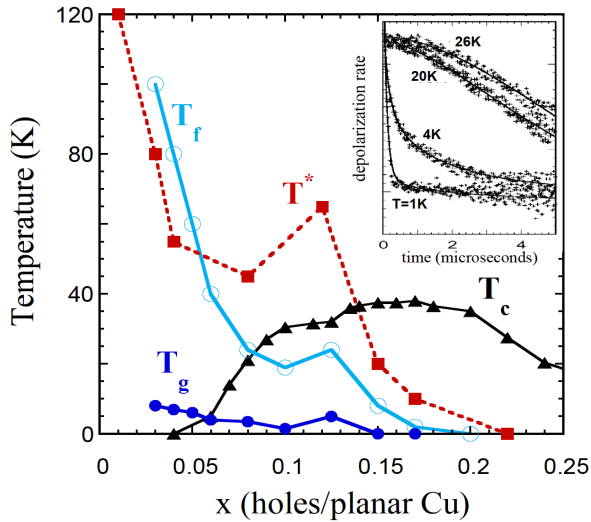}
\end{center}
\caption{Doping dependence of the superconducting transition temperature
$T_{c}$, the onset of slowing down of electronic moments, $T_{f,}$ before
being frozen at the glass temperature $T_{g}$. $T^{\ast}$ is a crossover from
metallic-like resisitivity to insulating-like. The inset depicts $\mu SR$ spectra
for $x=0.08$ at different temperatures, showing a deviation from a Gaussian
signalling the entrance of fluctuating electronic moments into the $\mu SR$
frequency window at approximately 25K (=$T_f$) and their eventual freezing below 4K (=$T_g$).}%
\label{fig2}%
\end{figure}From these studies we may conclude that superconductivity coexists
with glassiness on a microscopic scale throughout the bulk of the
material. This behavior is not limited to LSCO - although the
latter has been investigated most extensively. Similar results
have been observed in the \textquotedblleft
stripe-compound\textquotedblright\ Nd-LSCO, pure and Y doped
Bi-Sr-Ca-Cu-O, pure or Ca doped Y-Ba-Cu-O and more recently
Ca-Na-Cu-O-Cl
\cite{niedermayer-prl98,panagopoulos-ssc03,matsuda-prb02,sanna-lanl04,kohsaka-prl04,panagopoulos-prb02,kanigel-prl02}

\emph{Spin-charge correlations around the SC dome.} To further
examine the region where $T_{c}>0$, we re-analyzed early
measurements in high magnetic fields \cite{boebinger-prl96} where
bulk superconductivity was suppressed, revealing information about
low-$T$ charge transport in the normal phase. A striking
similarity is observed between the doping dependence of the spin
freezing and the resistivity minima all the way to the overdoped
region (Fig. 3). Both $T_{g}$ and $T^{\ast}$ decrease upon
doping,\thinspace\ except for an increase around $x=1/8$ (Fig. 3),
which is thought to reflect stripe pinning \cite{kivelson-rmp03},
or some other form of commensurate charge ordering
\cite{anderson-lanl04}. Furthermore, the doping dependence of the
resistivity minima closely tracks the onset of the slowing down of
spin fluctuations ($T=T_{f}$), before they freeze into a glassy
state ($T=T_{g}$). This observation is consistent with theoretical
predictions \cite{pastor-prl99} that the effective disorder
\[
W_{eff}=\left[  W^{2}+V^{2}q_{EA}\right]  ^{1/2}%
\]
seen by the charge carriers can be strongly enhanced by static and
dynamic fluctuations associated with glassy ordering. Here, $W$ is
the energy scale of the impurity potential, $V$ measures the
electron-electron interactions, and $q_{EA}$ is the frozen order
parameter fluctuation in the glassy phase \cite{pastor-prl99}.
Assuming that the self-generated randomness dominates the impurity
potential (i.e. $W^{2}\ll V^{2}q_{EA}$), this mechanism also
explains the correlations between the doping dependence of $T_{g}$
and $T^{\ast}$. This is true since one expects $T_{g}(x)\sim\left[
q_{EA}(x)\right]  ^{1/2}$, and the resistivity crossover scale
$T^{\ast}$ should be set by the effective disorder $W_{eff}$.
\vspace{12pt}

\emph{Intermediate conducting glass phase. }The correlation
between conductivity and glassiness indicates that for $x<x_{sc}$
($=0.05$) we are dealing with a strongly localized\ insulator
displaying hopping transport at $T<T^{\ast}$. Here, the number of
free carriers can be expected to vanish at $T=0$, in agreement
with recent Hall-effect measurements \cite{balakirev-nature03}. On
the other hand, for $0.05<x<0.20$ the number of carriers is found
to be finite \cite{balakirev-nature03}, suggesting an itinerant
system even in the normal phase. In this regime, DC transport has
a much weaker \cite{boebinger-prl96} (although still
insulating-like) temperature dependence. However, the observed
$\log T$ resistivity upturn in this region has been shown
\cite{boebinger-prl96} to be inconsistent with conventional
localization/interaction corrections which could indicate an
insulating ground state. Instead, estimates \cite{efetov-prl03}
reveal this behavior to be consistent with that expected for
metallic droplet charging/tunnelling processes, as seen in quantum
dots and granular metals \cite{efetov-prl03}. These results
suggest that in this regime HTS are inhomogeneous metals,
where conducting droplets connect throughout the sample, and a
metal-insulator transition in the normal phase happens
\textit{exactly} at $x=x_{sc}$. At lower densities the conducting
droplets remain isolated, and the material may be viewed as an
insulating cluster or stripe glass. As carrier concentration
increases they connect and the carriers are free to move
throughout the sample, forming filaments or "rivers". This is, in
fact, the point where  free carriers emerge in Hall-effect data
\cite{balakirev-nature03} and phase coherent bulk
superconductivity arises at $x>x_{sc}$. This observation indicates
that it is the \textit{inhomogeneous} nature of the underdoped
glassy region which controls and limits the extent of the
superconducting phase at low doping.

Therefore, based on evidence for charge retardation, freezing, and
uniformly distributed electronic heterogeneity in the form of
glassy stripes or droplets, we propose that in the interval
$x_{sc}<x<x_{opt}$ a novel intermediate phase arises in the form
of a bad metal. The emergence of such an intermediate conducting
glass phase separating a conventional metal and a glassy insulator
has, in fact, been predicted in recent theoretical work
\cite{mitglass-prl03}.

\emph{Quantum glass transition.} We now ask whether a true quantum critical
point (QCP) separates the glassy non-Fermi liquid and the metallic-Fermi
liquid-like regimes. We need an experiment where one may gradually increase
the amount of disorder, enhance short-range correlations, suppress
superconductivity, and fully expose the glassy ground state. These conditions
are met by $Zn^{2+}$ doping
\cite{panagopoulos-ssc03,panagopoulos-prb02,panagopoulos-prb04}. Figure 4
depicts characteristic data for La$_{2-x}$Sr$_{x}$Cu$_{0.95}$Zn$_{0.05}$%
O$_{4}$ with the normal state exposed ($T_{c}$=0) across the $T$-$x$ phase
diagram. The similar doping dependence of $T_{g}$ for pure and up
to 5\% Zn doped samples
\cite{panagopoulos-ssc03,panagopoulos-prb02,panagopoulos-prb04}
indicates that regardless of a sample being pure, disordered
(Zn-doped), superconducting or not we obtain universal behavior: A
set of glassy phase transitions, enhanced near $x=1/8$, and ending
at the same doping, supporting the presence of a quantum glass
transition insensitive to the amount of disorder. These results
strongly suggest that glassiness is not driven by impurities but
is predominantly self-generated, consistent with those theoretical
scenarios that predict phase separation \cite{kivelson-rmp03} at
low doping. Coulomb interactions, however, enforce charge
neutrality and prevent \cite{kivelson-rmp03} global phase
separation; instead, the carriers are expected
\cite{schmalian-prl00} to segregate into nano-scale domains -
 to form a stripe/cluster glass \cite{schmalian-prl00}. As
quantum fluctuations increase upon doping
\cite{pastor-prl99,mitglass-prl03}, such a glassy phase should be
eventually suppressed at a quantum critical point, which in LSCO
emerges around $x=x_{opt}\approx 0.2$.  Remarkable independent
evidence that a QCP is found precisely at $x=x_{opt}$ is provided
by the observation of a sharp change in the superfluid
density $n_{s}(0)\sim1/\lambda_{ab}^{2}(0)$ (where $\lambda
_{ab}(0)$ is the absolute value of the in-plane penetration
depth). At $x>x_{opt}$, $n_{s}(0)$ is mainly doping independent
(Fig. 4), while the $T$-dependence is in good agreement with the
BCS weak-coupling $d$-wave prediction \cite{panagopoulos-ssc03}.
At dopings below the quantum glass transition $n_{s}(0)$ is
rapidly suppressed (note the enhanced depletion near $x=1/8$
precisely where $T_{g}$ and $T^{*}$ are enhanced) and there is a
marked departure of $n_{s}(T)$ from the canonical weak coupling
curve \cite{panagopoulos-ssc03}. Similar behavior has been
observed in other HTS and in the $c$-axis component
\cite{panagopoulos-ssc03}. The penetration depth data show that
the onset of quasi-static magnetic and charge correlations
coincides with an abrupt change in the superconducting ground
state. In addition, a crossover temperature $T_m$ at $x>x_{opt}$
separating marginal Fermi liquid transport at $T> T_m$ from more
conventional metallic behavior at $T<T_m$ also seems to drop
\cite{naqib-physicaC03} to very small values around optimum doping
(see Fig. 1). At $x>x_{opt}$ the ground state becomes metallic and
homogeneous, with no evidence for glassiness or other form of
nano-scale heterogeneity
\cite{panagopoulos-ssc03,panagopoulos-prb02,boebinger-prl96,balakirev-nature03,panagopoulos-prb04,davis-lanl04}.
All these results provide strong evidence of a sharp change
in ground state properties at $x=x_{opt}$, and the emergence of
vanishing temperature scales as this point is approached - just
as one expects at a QCP. Let us note, the extent of the region between $x_{sc}$ and $x_{opt}$
is material dependent and expected to vary across the HTS families.
\begin{figure}[ptb]
\begin{center}
\includegraphics[
height=3.4529in,
width=3.208in
]{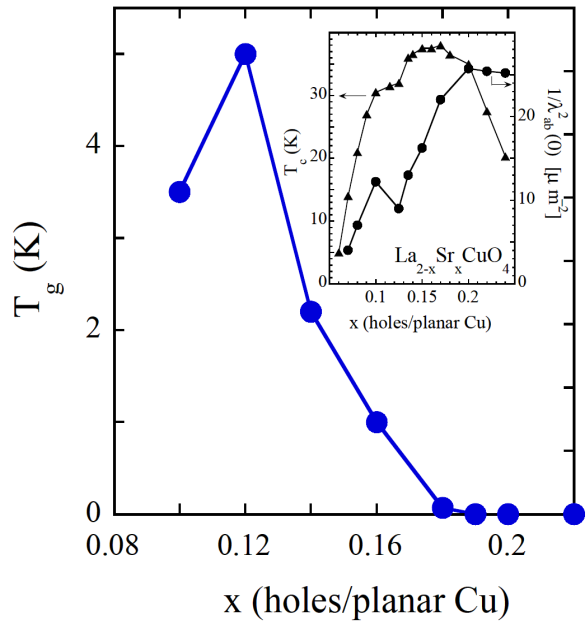}
\end{center}
\caption{Doping dependence of the glass transition temperature $T_{g}$ for
LSCO doped with 5\% Zn. The inset shows data for the superconducting
transition temperature $T_{c}$ and superfluid density $n_{s}(0)\sim1/\lambda_{ab}^{2}(0)$
for the pure LSCO system, indicating the transition in the superconducting ground state precisely at the
concentration where glassiness vanishes. }%
\label{fig3}%
\end{figure}

In summary, we have identified three\ distinct doping regimes: (1)
$x<x_{sc}$; (2) $x_{sc}<x<x_{opt}$; and (3) $\ x>x_{opt}$,
separated by two critical points: a quantum glass transition at
$x=x_{opt}$ and a normal state metal-insulator
 transition at $x=x_{sc}$ within the glassy phase. On
this basis we propose the behavior of HTS could bear resemblance to other
materials close to disorder-driven metal-insulator transitions, where
electronic heterogeneity and self-generated glassiness arise with the first
added holes - a mechanism that may potentially explain many puzzling features
of cuprate superconductors.

CP acknowledges earlier collaboration with M. Kodama, E.
Liarokapis, T. Nishizaki, N.J. Owen and T. Sasagawa. We thank L.
Gor'kov, E. Manousakis, S. Sachdev, and J. Schmalian for helpful
discussions. The work in Cambridge was supported by The Royal
Society and the ISIS Rutherford Appleton Laboratory. The work at
FSU was supported through grant NSF-0234215.


\newcommand{\noopsort}[1]{} \newcommand{\printfirst}[2]{#1}
  \newcommand{\singleletter}[1]{#1} \newcommand{\switchargs}[2]{#2#1}

\end{document}